\documentclass[preprint]{revtex4}
\usepackage{graphics}
\begin{document}

\title{Power-law exponent of the Bouchaud-M\'ezard model on regular random
 network}
\author{Takashi Ichinomiya}
\affiliation{Department of Biomedical Informatics, Gifu University
 School of Medicine, Yanagido 1-1, Gifu 501-1194, Gifu, Japan}
\date{\today}
\email{tk1miya@gifu-u.ac.jp}
\begin{abstract}
 We study the Bouchaud-M\'ezard model on a regular random  network. By
 assuming  adiabaticity and independency, and utilizing the generalized
 central limit theorem and the Tauberian theorem, we derive an
 equation that determines the exponent of the probability distribution
 function of the wealth as $x\rightarrow \infty$. The analysis shows that
 the exponent can be smaller than 2, while a mean-field analysis
 always gives the exponent as being larger than 2. The results of our
 analysis are shown to be good agreement  with those of the numerical
 simulations.
\end{abstract}

\maketitle

\section{Introduction}
Since  Pareto's revelation that the distribution of wealth follows a
 power-law\cite{Pareto}, many researchers have proposed various models
 to explain this phenomenon.
The simplest of these is the
Bouchaud-M\'ezard (BM) model, which is described by the following
stochastic differential equation\cite{Bouchaud2000};
\begin{equation}
 dx_i = J\sum_{j=1}^{N} a_{ij}(x_j-x_i) dt +\sqrt{2}{\sigma} x_i d\xi_i,\label{085140_10Apr13}
\end{equation}
where $N$, $x_i$, $J$, $a_{ij}$, $\sigma$, and $\xi_i$ are the total
number of agents, the wealth of agent
$i$, a diffusion constant,
an adjacent matrix, the noise strength, and normal Brownian motion,
 respectively. 
 Bouchaud and M\'ezard\cite{Bouchaud2000} analyzed this
 model for a globally coupled network, i. e.,  $a_{ij}=1$ for all $i\ne
 j$, and  found that, although $x$ does not have a  stationary
 distribution,  the normalized
 wealth $x/\langle x \rangle$ does and its probability distribution
 function (PDF) $\rho(x)$ is
 given by 
 \begin{equation}
 \rho(x)=\frac{\mu_N }{\Gamma(1+\mu_N)} \exp\left[-\frac{\mu_N}{x}\right]x^{-2-\mu_N},
 \end{equation}
where $\mu_N=JN/\sigma^2$ (here and in the following, we denote the normalized wealth by $x$). The PDF obeys the power law  
$\rho(x)\propto x^{-\beta}$ as $x\rightarrow\infty$, where the exponent
is $\beta=2+\mu_N$.
 
 Real economic networks, however, are not  globally coupled networks,
 and so the original BM model is extended to that on a complex
 network.
 However, it has been shown that the properties of the BM model on a complex
 network differ from those on a globally coupled network.
 Bouchaud and M\'ezard\cite{Bouchaud2000} also reported that, while the
 mean-field analysis
 always gives an exponent larger than 2, a smaller exponent can be seen
 in  numerical simulations on a regular random network.
 This result seems plausible because the real
 wealth distribution often shows an exponent that is  smaller than 2\cite{Sinha2006}.
 So far, a number of papers have been published on this system.
 For example, Garlaschelli and Loffredo proposed to fit the distribution
 by $\rho(x) \propto \exp(-\alpha/x)x^{-\beta}$\cite{Garlaschelli2004}. Ma {\it et al.}
 proposed another assumption that the  PDF is given generalized inverse
 Gamma distribution\cite{Ma2013}. In these papers, the exponent $\beta$
 was obtained by fitting the result of numerical simulations, however,
 no one had succeeded to obtain $\beta$ analytically. 

 We recently obtained the 
  wealth distribution of the BM model on the complex network
  analytically in the case  of $\beta > 3$\cite{Ichinomiya2012-1, Ichinomiya2012-2}. 
However, these analysis cannot be applied when
$\beta < 3$, because we derive the result
  using the central limit theorem. We can apply the central limit
  theorem only when the variance is finite, therefore this theorem is
  not applicable  when $\beta < 3$. 

Here, a study of the power-law tail of the BM model on a regular random
network when $\beta < 3$ is presented.
The applications of the generalized central limit theorem and the
Tauberian theorem are a key part of this theory. 
Using these theorems, we derive two equations that relate the PDF of the
wealth to that of the average of wealth around each node. 
These two equations lead us to a simple equation that determines the
exponent of the stationary wealth distribution.
 The result shows that the exponent can be smaller
 than $2$, which implies that ``wealth condensation''(a small number
 of agents control the greater part of the wealth) can occur.

We first develop the theory for the wealth distribution  of the BM model
 on a regular random network in  Sec.  \ref{084857_10Apr13}, and compare 
 the numerical simulation results with our analysis in
 Sec. \ref{132621_9Apr13}. 
We summarize the result and discuss  the
generalization of our theory and remaining problems in Sec. \ref{152701_16Apr13}.

\section{Theory}
\label{084857_10Apr13}

 Following the theory presented in \cite{Ichinomiya2012-1,
 Ichinomiya2012-2}, 
we make ``adiabatic''
 and ``independent'' assumptions. We assume that the PDF of the wealth on
 each node is independent, allowing the  total PDF
 $\rho(x_1, x_2,\cdots x_N)$ to be decomposed as
 $\rho(x_1, \cdots, x_N) =\prod_i \rho_i(x_i)$.
 In the case of a regular random network in which all nodes
 have degree $k$, $\rho_i(x_i)$ can be assumed to be independent of the
 index $i$, i. e., $\rho_i(x_i) = \rho(x_i)$. We also assume
 ``adiabaticity'', that is, $x_i$ varies much
 faster than the ``local mean-field'' $\bar x_i = \frac{1}{k}\sum_j a_{ij}x_j$. 

 If  $\bar x_i$ is constant, the conditional PDF of $x_i$, $\rho(x_i|\bar x_i)$, is given by the stationary
 solution of the following Fokker-Planck equation:
\begin{equation}
  \frac{\partial \rho}{\partial t} = -\frac{\partial}{\partial
  x}[k(\bar{x}_i-x_i)\rho_i] +\sigma^2 \frac{\partial^2}{\partial x_i^2}
  [x_i^2 \rho_i].\label{162421_3Apr13}
\end{equation}
 It can be shown that 
\begin{equation}
 \rho(x_i|\bar x_i) = C(\bar x_i)\exp(-\mu_k \bar{x}_i/x_i)x_i^{-2-\mu_k}\label{144129_18Mar13}
\end{equation}
 is the stationary solution of this equation, where
\begin{equation}
 C(\bar x)=(\mu_k \bar x)^{1+\mu_k}/\Gamma(1+\mu_k),\label{092714_2Apr13}
\end{equation}
  and $\mu_k=Jk/\sigma^2$.
Under the adiabatic assumption, $\bar x_i$ changes at a much slower rate
than $x_i$, and so $\rho(x)$ can be written as 
\begin{equation}
 \rho(x) \sim \int_0^{\infty} d\bar{x} \rho(x|\bar x )P(\bar{x}),\label{144057_18Mar13} 
\end{equation}
 where $P(\bar x)$ is the distribution of the average of the wealth around
 node $i$.

To calculate the exponent as $x\rightarrow \infty$,  we assume
 that $\beta<2+\mu_k$ and $\beta < 3$. This assumption is consistent with the
 numerical simulations results for small $\mu_k$  that are presented  in
 Sec. \ref{132621_9Apr13}.
 Under this assumption, we can obtain two equations that relate the
 tail of $\rho(x)$ to $P(\bar{x})$ as $x\rightarrow \infty$.

The first of these equations is derived by applying the
 generalized  central limit theorem\cite{Feller1966}.
Because the standard deviation of $x$ diverges, we can assume that
$1-\int_0^X dx \rho(x) \sim C^\prime X^{-\alpha}$ as $x\rightarrow
 \infty$ and  $\alpha \le 2$. 
Under  the independent assumption, we can apply the generalized central
limit theorem, and the distribution of $p_k \sum_j a_{ij}(x_j-1)$ can be
approximated  by the stable distribution $S(x;\alpha,1,1)$ for large $k$,
 where $p_k =[2\Gamma(\alpha)\sin(\alpha\pi/2)/(\pi C^{\prime})]^{1/\alpha}
 k^{-1/\alpha}$, and $S(x;\alpha,\gamma, \beta)$ is the stable distribution
 function whose characteristic function is given by $\omega(k)
 =\exp[-\gamma^\alpha|k|^\alpha + i \beta \gamma^\alpha k |k|^{\alpha-1} \tan \frac{\pi
 \alpha}{2}]$ for $\alpha \ne 1$  and $\omega(k) =
 \exp[-\gamma |k|-i \beta \gamma \frac{2}{\pi}k\log|k|]$ for $\alpha =1$. 
In both cases,  the tail of $s(x;\alpha,1,1)$, which is the PDF of the
 stable
 distribution
  $S(x;\alpha,1,1)$, is given by
\begin{equation}
  s(x;\alpha,1,1) \sim 2  \sin(\pi \alpha/2)\Gamma(1+\alpha) x^{-1-\alpha}/\pi\label{161104_2Apr13}
\end{equation}
 for $x\rightarrow \infty$\cite{Nolan2013}. Here, we note that the tail
 of PDF of 
 $\sum_j a_{ij}(x_j-1)$ and $\sum_{j} a_{ij}x_j$ is the same for $x\rightarrow
 \infty$. 

By assuming that $\rho(x)\sim C x^{-\beta}$ and
$P(\bar x)\sim D {\bar x}^{-\beta}$ as $x, \bar x \rightarrow \infty$, we obtain the following
equations from Eq.(\ref{161104_2Apr13}) and the expression for $p_k$;
\begin{equation}
 \beta = \alpha+1,
\end{equation}
\begin{equation}
 C = \alpha C^{\prime},
\end{equation}
and 
\begin{equation}
 D = \alpha  k^{1-\alpha} C^{\prime}.
\end{equation}
These equations  lead to 
\begin{equation}
 D=k^{2-\beta}C\label{095908_19Mar13},
\end{equation}
which is the first equation we use to determine the exponent $\beta$.

The second equation is obtained using the Tauberian theorem.
We introduce the Laplace transformation of $f(x)$,
\begin{equation}
 \hat f(s) = \int_0^\infty dx e^{-s x} f(x)  .
\end{equation}
The Tauberian theorem states that  $\hat f(s)\sim Cs^{-\gamma-1}$ as
 $s\rightarrow 0$ is equivalent to $f(x) \sim \frac{C}{\Gamma(\gamma+1)}
x^\gamma$ as $x\rightarrow \infty$, provided that $\gamma > -1$.

 Eqs. (\ref{144129_18Mar13}), (\ref{092714_2Apr13}) and (\ref{144057_18Mar13}) 
 indicate that $\rho(x)$  can be obtained from $\widehat{CP}(s)$, the Laplace transformation of
 $C(\bar x) P(\bar x)$, as
\begin{equation}
 \rho(x) = x^{-2-\mu_d} \widehat{CP}(\frac{\mu_k}{x}) .\label{131829_10Apr13}
\end{equation}
Using  the fact that $C(\bar x) P(\bar x) \propto \bar{x}^{1+\mu_k-\beta}$ as
 $x\rightarrow \infty$ and the assumption $\beta <2+\mu_k$, we can apply
 the Tauberian theorem to calculate $\widehat{CP}(\frac{\mu_k}{x})$
 as $x\rightarrow \infty$;
\begin{equation}
 \rho (x) = D x^{-2-\mu_k}\frac{\mu_k^{1+\mu_k}}{\Gamma(1+\mu_k)}\Gamma(2+\mu_k-\beta)
  \left(\frac{x}{\mu_k}\right)^{2+\mu_k-\beta} = D  x^{-\beta} \mu_k^{\beta-1}\frac{\Gamma(2+\mu_k-\beta)}{\Gamma(1+\mu_k)}\label{135524_17Apr13}
\end{equation}
for $\mu_k/x \rightarrow 0$. Therefore, we obtain the second equation,
\begin{equation}
 C  = D
\frac{\mu_k^{\beta-1}\Gamma(2+\mu_k -\beta)}{\Gamma(1+\mu_k)}\label{095922_19Mar13}.
\end{equation}

Combining Eqs.(\ref{095908_19Mar13}) and (\ref{095922_19Mar13}), we can
 determine the exponent $\beta$ from
\begin{equation}
 \frac{\Gamma(2+\mu_k-\beta)\mu_k^{\beta-1}}{\Gamma(1+\mu_k)} = k^{\beta-2}\label{124304_19Mar13}.
\end{equation}

There are a few things to note about this equation. First,
this equation is always satisfied when $\beta=2$; 
however, it seems counterintuitive that $\beta$ does not depend on the
parameter $J$, and so we take the solution that satisfies $\beta\ne 2$ as
the ``real'' distribution.
Second, it should also be noted that in general this equation cannot be
solved analytically, but 
we can obtain $\mu_k$ in some special cases, e. g., when
$\beta\rightarrow 3$.
In this case,
Eq.(\ref{124304_19Mar13}) gives 
\begin{equation}
 k=\mu_k^2\frac{\Gamma(\mu_k -1)}{\Gamma(\mu_k +1)}=\frac{\mu_k}{\mu_k-1}.
\end{equation}
The solution to this  equation,
\begin{equation}
 \mu_k = k/(k-1),
\end{equation}
 implies that the variance of the wealth becomes finite  when  $ J > J_c =
 \sigma^2/(k-1)$. This result is consistent with that in our previous
 paper\cite{Ichinomiya2012-1}, in which we investigated the
wealth distribution when $\beta > 3$.
Third, we note also that in the limit $k\rightarrow \infty$, we
 obtain $\beta = 2+\mu_k$ from Eq.(\ref{124304_19Mar13}), which is the
 solution of the mean-field analysis; as $k\rightarrow \infty$, the
 right hand side of Eq.(\ref{124304_19Mar13}) diverges for $\beta >2$,
 and is 0 for $\beta<2$.
 From the fact that $\Gamma(x)$ has no zero point on the real
 axis and  diverges at $x=0$, we conclude that $\beta\rightarrow 2+\mu_k$ as $k$
 increases, which is consistent with the results of the mean-field analysis.
 In a final remark about Eq.(\ref{124304_19Mar13}), we see that there is
 a solution $\beta \le 2$ for small
$\mu_k$, as we show in Sec. \ref{132621_9Apr13}. This appears to be
 paradoxical, because the average of the normalized wealth  must be 1,
 while it diverges if $\beta \le 2$.
 This paradox is a result of the independent assumption; because the sum
 of $x_i$ is conserved, $x_i$ and $\bar x_i$ are related through  $x_i +
 k \bar x_i < N$. In the  case of $\beta > 2$,  this restriction is
 negligible, because $x_i=o(N)$ for all nodes.
 However, in the case of $\beta < 2$, the wealth of the
 richest node $x_M$ becomes $x_M=O(N)$, and the correlation is not negligible.
 Therefore for   $x=O(N)$ the distribution will differ from that 
 derived here; our theory can only be applied in in the region $x \ll N$.

 Before concluding this section, we make a remark on the
 generalization of our method to a general complex network.
 We assume again  adiabaticity and independency,
 and we also assume  that the PDF of $x_i$ and that of the local
 mean-field $\bar x_i$ at
 node $i$ are given by $\rho_i(x)\sim C_i x^{-\beta}$
 and $P_i(\bar x) \sim D_i \bar{x}^{-\beta}$ for $x,\bar{x} \rightarrow
 \infty$, respectively.
 Under these assumptions, all $\rho_i(x)$ belong to the same domain of
attraction of a stable distribution function. In this case, the
distribution of the sum $\sum_j a_{ij} x_j$ can be
 obtained from the generalized central limit theorem\cite{Otiniano2012},
which leads to
\begin{equation}
 D_i = k_i^{1-\beta} \sum_j a_{ij}C_j,\label{144503_4Apr13}
\end{equation}
where $k_i$ is the degree of node $i$.
By also using the Tauberian theorem we also get
\begin{equation}
 C_i = D_i \frac{\mu_{k_i}^{\beta-1} \Gamma(2+\mu_{k_i}-\beta)}{\Gamma(1+\mu_{k_i})},\label{144511_4Apr13}
\end{equation}
and by eliminating $D_i$ from Eqs.(\ref{144503_4Apr13}) and
 (\ref{144511_4Apr13})
we obtain
\begin{equation}
C_i =
  \mu_{k_i}^{\beta-1}\frac{\Gamma(2+\mu_{k_i}-\beta)}{\Gamma(1+\mu_{k_i})}k_i^{1-\beta}\sum_j a_{ij}C_j,
\end{equation}
which can be rewritten as 
\begin{equation}
 G C =  A C,\label{093420_12Apr13}
\end{equation}
where
\begin{equation}
(G)_{ij} = \delta_{ij}\frac{k_i^{\beta-1}\Gamma(1+\mu_{k_i})}{\mu_{k_i}^{\beta-1}
 \Gamma(2+\mu_{k_i}-\beta)}, (A)_{ij} = a_{ij}, (C)_i = C_i\label{152942_4Apr13}.
\end{equation} 
Therefore we require $\beta$ to satisfy the condition $\mbox{det}(G-A)
=0$, and we also require the eigenvector of $G-A$, whose corresponding
eigenvalue is zero, to be non-negative, because $C_i \ge 0$ for all $i$.
In the case of a regular random graph, we can find the solution easily;
first, $G$ is proportional to the identity matrix, and the condition 
$\mbox{det}(G-A)=0$ means that the diagonal part of $G$ is equal to
an eigenvalue of $A$. Second, $A$ is a non-negative matrix and  we can
 apply the Frobenius theorem to find that $A$ has a  non-negative
 eigenvector $(1,1,\cdots,1)$ and a corresponding eigenvalue  $k$.
 From these conditions, we can easily obtain
Eq.(\ref{124304_19Mar13}).
 In the case of a general complex network, it is
much more difficult to obtain $\beta$;   $G$ is not
 proportional to the identity matrix, and we need to solve
 $\mbox{det}(G-A) = 0$
 numerically.
 Moreover, we cannot apply the
 Frobenius theorem to obtain non-negative eigenvectors, because $G-A$ is
 not  a non-negative matrix.  In this case, we will need to follow the
 following procedure to obtain $\beta$: First, we find the set of $\beta$'s that
satisfy $\mbox{det}(G-A) =0$. Second, we  select those $\beta$ values that give
a  non-negative eigenvector of $G-A$ whose eigenvalue is zero.  We note
that it is difficult  to carry out this procedure for a large
 complex network. In the first
 step, we will find a large number of  $\beta$'s that satisfy $\mbox{det}(G-A)=0$.
 To estimate the number of solutions of $\mbox{det}(G-A)=0$, we consider
 the case of the regular random graph. In this case,
 the number of $\beta$'s that  satisfy this condition is equal  to the
 number of eigenvalues of $A$. Therefore we expect that the number of $\beta$'s
 that satisfy this condition will be  $O(N)$ for a large  complex
 network. From this set of  $\beta$'s, we need to choose those  that
 have  non-negative eigenvectors. We cannot apply the Frobenius theorem
 in the case of general complex network, and we need to calculate the
 corresponding eigenvectors for each $\beta$ obtained in the first step. 
 And even if we succeeded in doing this, we also need to select one 
 appropriate $\beta$ from this set. In the case of a regular random network, we always
 have the solution $\beta=2$, which is rejected because it is
 counterintuitive. However, it is not known whether we can always
 distinguish appropriate $\beta$ values from inappropriate values simply by
 intuition. Further work will be necessary to extend this theory to the
 case of a general complex network.

\section{Numerical Simulation}
\label{132621_9Apr13}

Numerical simulations of the BM model on a regular random graph
 were carried out for $N=5000$ nodes and repeated 10 times using the
 Euler-Maruyama algorithm
 with an initial condition $x_i=1$ for each parameter. 
 The distribution at $t=2000$ was taken as the stationary distribution,
 and here we set $\sigma^2 = 1$.

\begin{figure}[t]
 \resizebox{.45\textwidth}{!}{\includegraphics{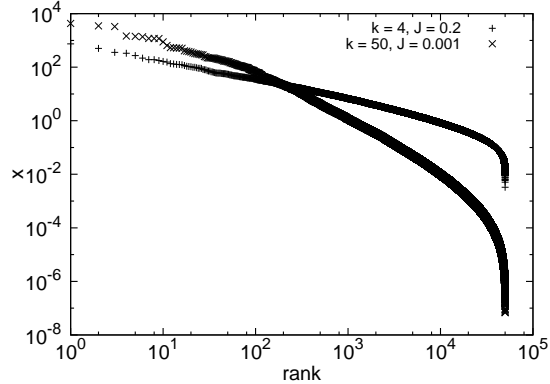}}
\caption{Rank plot of the wealth $x$ obtained from numerical simulations for
 the cases of $k = 4, J = 0.2$ and $k = 50, J = 0.001$.}
\label{134055_8Apr13}
\end{figure}

 To obtain the exponent at large $x$, we construct a rank plot of $x$; it has
 been  shown previously that if  $\rho(x)\propto x^{-\beta}$ for $x
 \rightarrow \infty$, then the wealth of the $l$-th richest agent is
 approximately proportional to $l^{\frac{1}{\beta-1}}$\cite{Redner1998}.
 A rank plot of the wealth for $k=4, J=0.2$ and $k=50, J=0.001$ is shown
 in Fig.\ref{134055_8Apr13}. 
 The distribution shows  clear power-law behavior in  the case of
 $k=4, J=0.2$, but not  in the case of  $k=50, J=0.001$. 
Although the exponent $\beta$ seems to be constant at
 ranks of  $100 < i < 1000$, it increases for smaller
 $i$. This observation is consistent with the discussion  at the end
 of Sec. \ref{084857_10Apr13}. For $k=50$ and  $J=0.001$, 76 agents
 have a wealth larger than 100, and so we cannot  assume that $x_i \ll N$
 in this region.
 From these observations,  we  estimate the
 exponent as $\beta = 1+ 1/\log_{10} (x_{100}/x_{1000})$, where $x_i$
 represents the wealth at rank $i$.

\begin{figure}[t]
 \resizebox{.45\textwidth}{!}{\includegraphics{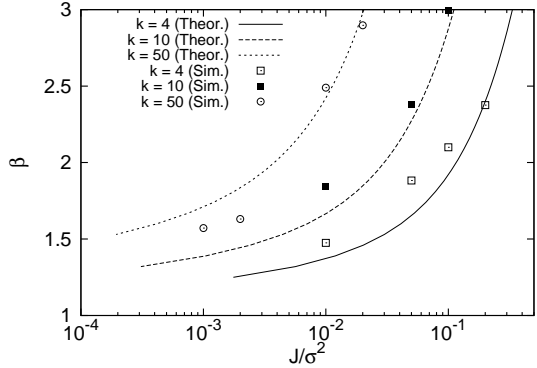}}
\caption{Values  of the exponents obtained from our theory (lines) and the numerical
 simulation (symbols). Results are shown for  $k = 4, 10$ and
 50.}
\label{092544_3Apr13}
\end{figure}

 Fig.\ref{092544_3Apr13} compares the exponents obtained in the  numerical
 simulation for several parameters with those  of the  theory presented
 in Sec. \ref{084857_10Apr13}. Good agreement is observed, especially in
 case of $\beta >2$, but at lower $\beta$ values there is a small discrepancy
 between the theoretical and simulation values.
 We suspect that this discrepancy is caused by the errors in the
 estimation of the exponent.
 As shown in Fig.\ref{134055_8Apr13}, the  wealth distribution does not
 exhibit a power-law in the case of  $\beta \le 2$  due to the
 correlation for large $x$.
 We use  data that have a  rank of larger than 100 to avoid the
 effect of this correlation; however, the correlation will  still
 remain. We also note that $x_{1000}$ becomes too small to assume $\mu_k/x
 \ll 1$, which  is  an assumption used in the derivation of Eq. (\ref{131829_10Apr13}).
 If  another method  is used to estimate the exponent, such
 as $\beta = 1+1/\log_{5}(x_{100}/x_{500})$, then we will obtain a
 slightly different exponent. 
 Despite  this problem,
 the difference in the value of the exponent obtained by our theory and
 that obtained from the  numerical simulations
 is  less than $\sim 0.1$, and we can conclude that
 our theory gives a good estimation of the  exponent even in the case of
 $\beta < 2$.

\section{Discussion and Conclusion}
\label{152701_16Apr13}
 In summary, we have analyzed the stationary PDF of wealth in
 the BM model on a regular random network. Using
 the generalized central limit theorem and the Tauberian theorem, we
 obtained an equation for the  tail exponent $\beta$. We found that
 at small coupling the exponent $\beta$ becomes smaller than 2, and the
 results of this analysis were in good agreement with those of the
 numerical simulations. 

 This work has raised a number of points that require further investigation.

 The first important point to be addressed is the generalization of our
 theory to the BM model on a complex network. As discussed at the end of
Sec. \ref{084857_10Apr13}, we need to obtain a value of $\beta$ that
 satisfies the condition that
 the matrix $G-A$  has  a zero eigenvalue and that the 
corresponding eigenvector is non-negative. In  practice, it is difficult
 to  find a value of 
$\beta$ that satisfies these conditions. In the case of a regular random
graph, we can  assume that $C_i = C$ for all
$i$. This assumption cannot be applied if the network is heterogeneous.
Numerical simulations on a heterogeneous network may suggest  other
assumptions on $C_i$, which would assist in calculating $\beta$.

The second important point is an  estimation of  the effects of correlation.
Medo\cite{Medo2008}  showed  that there is a  spatial correlation in
 the BM model on complex networks. We have
already reported that such a correlation is not negligible in the case of
a Watts-Strogatz network\cite{Ichinomiya2012-2}.
In this paper, we found that  another kind of correlation emerges from
the wealth conservation. This correlation  can occur in any complex
network and is not negligible especially for $\beta < 2$. Further
studies for a  more accurate  estimation of  the effects caused by 
 this correlation are needed.

The final point is  the  unification  of this work with our previous
 work for $\beta >3$. In the case of $\beta > 3$, we
can approximate $P(\bar x)$ by a Gaussian distribution using the central
limit theorem.
Unfortunately, we cannot apply the Tauberian theorem in this case.
 This theorem is applicable only when
$C(\bar x) P(\bar x) \sim \bar x^{-\gamma} L(\bar x)$ for $\bar x
\rightarrow \infty$, where $L(\bar x)$ is a
slowly varying function of $\bar x$, i.e., $L( t \bar x )/L(\bar x
)\rightarrow 1$ as $t\rightarrow \infty$ for all $\bar x>0$. If $P(\bar
 x)$ is a  Gaussian, then there is no $L(\bar x)$ that satisfies
this condition, and so we cannot apply the Tauberian theorem. We will need
another approach to unify both cases into a single theory.

\begin{acknowledgments}
 The author thanks Hiroya Nakao, Satoru Morita, and Takaaki Aoki for
 helpful discussions.
\end{acknowledgments}

\end{document}